# Imaging of moving targets with multi-static SAR using an overcomplete dictionary

Ivana Stojanovic, *Member, IEEE,* and William C. Karl, *Senior Member, IEEE*

ECE Department, Boston University
8 St. Mary's St, Boston, MA 02215
E-mail: ivanas@bu.edu and wckarl@bu.edu

*Abstract*—**This paper presents a method for imaging of moving targets using multi-static SAR by treating the problem as one of spatial reflectivity signal inversion over an overcomplete dictionary of target velocities. Since SAR sensor returns can be related to the spatial frequency domain projections of the scattering field, we exploit insights from compressed sensing theory to show that moving targets can be effectively imaged with transmitters and receivers randomly dispersed in a multi-static geometry within a narrow forward cone around the scene of interest. Existing approaches to dealing with moving targets in SAR solve a coupled non-linear problem of target scattering and motion estimation typically through matched filtering. In contrast, by using an overcomplete dictionary approach we effectively linearize the forward model and solve the moving target problem as a larger, unified regularized inversion problem subject to sparsity constraints.**

## I. INTRODUCTION

Synthetic aperture radar (SAR) is a remote sensing system capable of producing high-resolution imagery of target scenes independent of time of day, distance, and weather. Conventional SAR radars are monostatic, with collocated transmit and receive antenna elements. These SAR sensors coherently process multiple, sequential observations of a scene under the assumption the scene is static. When the scene changes between these observations, as occurs when objects move, and these changes are ignored, blurring, defocus, and other artifacts are introduced into the reconstructed imagery. This is because the Doppler shift of moving objects are then determined not only by their geometric location but also by their velocity. Imaging of scenes with moving targets has gained increasing interest as the desire for persistent and urban sensing has grown.

Moving target localization has proven challenging in the case of single antenna conventional narrow-angle SAR utilizing conventional reconstruction methods, such as the polar format and the filtered-back-projection algorithms [1], [2], due to an inherent ambiguity in target geolocation and velocity. Consequentially, most techniques for imaging moving targets with conventional SAR aim at focusing and detecting smeared targets in SAR imagery [1], [2], [3], [4], [5]. In recent years, however, a number of techniques to handle moving objects explicitly have been developed. Space-time adaptive processing (STAP) [6] exploits multiple-phase center antennas to suppress clutter and produce a moving target indication image. Velocity synthetic aperture radar (VSAR) [7] exploits the velocity information contained in phases of a sequence of images formed at multiple receive antenna elements. Dual-speed SAR [8] has the radar platform move sequentially at two different velocities during the radar data collection time. Distributed antenna radars also have the potential to break the velocity-location ambiguity due to multiple phase centers of the antenna, while at the same time substituting spatial diversity for conventional bandwidth resources. Recent work on multi-static [9] and the related MIMO [10], [11], [12] radar with coherent processing has shown the potential for resolution improvement that can far exceed the limit suggested by conventional arguments based on the radar's waveform.

The conventional approach to resolving the moving target localization problem is to perform matched filter reconstruction at every pixel for every possible velocity hypothesis independently, yielding a large space-velocity cube [9], [4]. A target is then placed at the locations of maximal energy focus in the space-velocity cube. An approach based on the inversion of the forward operator is presented in [13], where a filltered-backprojection approach to imaging of moving targets takes the form of a weighted matched filter. These approaches require the solution of many large inversions and result in a large, somewhat ambiguous output.

In this paper we utilize an inverse problem formulation and insights from sparse signal representation and compressed sensing for effective imaging of dynamic environments using distributed antenna SAR sensor geometries. The non-linear problem of the coupled target localization and velocity estimation is linearized by construction of an overcomplete dictionary of velocity states. The resulting inversion leads to a non-convex optimization problem which we efficiently solve through convex relaxation. In contrast to the filtered-backprojection and the matched filtering approaches of [13] and [9], in our approach, velocity estimation is performed during the image formation process and all velocity hypothesis are evaluated jointly in a single optimization framework.

## II. OBSERVATION MODEL

We consider a multi-static system consisting of widely separated transmit and receive elements within a forward cone positioned at the center of a scene of interest. We assume that different transmitters send out probes in TDMA fashion, while the receive unit coherently processes signals of all receivers across all snapshots, i.e. pulses. The scene of interest is modeled by a set of point scatterers reflecting impinging electromagnetic waves isotropically to all receivers within the forward cone, thus, allowing for the coherent processing of all received signals. The reflection coefficient of the point scatterer is a complex number with an unknown amplitude and a random phase [14].

We introduce a coordinate system with the origin in the center of the area of interest and, for simplicity, model the scene as two dimensional. Fig. 1 illustrates this set up. The relative size of the scene is assumed to be small compared to distances from the origin of the coordinate system to all transmitter and receivers, such that transmit and receive angles would change negligibly if the coordinate origin moved to any point in the scene. Furthermore, we neglect signal propagation attenuation.

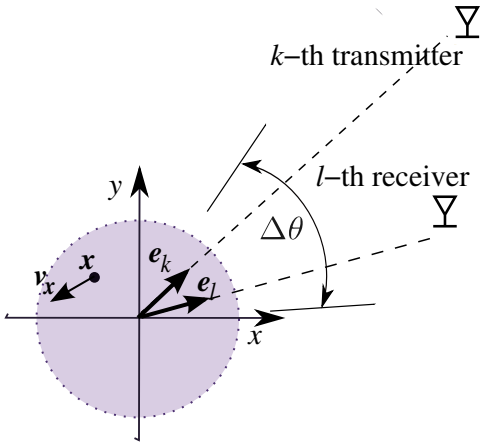

Fig. 1. Geometry of the $kl$-th transmit-receive pair with respect to the scene of interest. All transmit and receive pairs are restricted to lie within a forward cone of the angular extent $\Delta\theta$.

### A. Stationary scene model

The complex signal received by the $l$-th receiver for the excitation from the $k$-th transmitter reflected from a point scatterer at the spatial location $\mathbf{x} = [x, y]^T$ is given by

$$r_{kl}(t) = s(\mathbf{x})\, \gamma_k \left(t - \tau_{kl}(\mathbf{x}_o) - \tau_{kl}(\mathbf{x})\right),$$

where $s(\mathbf{x})$ is the reflectivity of the scatterer, $\gamma_k(t)$ is the transmitted waveform from the $k$-th transmitter, $\mathbf{x}_o = [0, 0]^T$ is the scene's origin and $\tau_{kl}(\mathbf{x}_o)$ is the sum of the signal propagation delay from the $k$-th transmitter to the scene's origin, $\tau_k(\mathbf{x}_o)$, and the propagation delay from the scene's origin to the $l$-th receiver, $\tau_l(\mathbf{x}_o)$, so that $\tau_{kl}(\mathbf{x}_o) \doteq \tau_k(\mathbf{x}_o) + \tau_l(\mathbf{x}_o)$. Under the far field assumption, the propagation delay from the scene's origin to the scatterer at the location $\mathbf{x}$ is given by:

$$\tau_{kl}(\mathbf{x}) \doteq \tau_k(\mathbf{x}) + \tau_l(\mathbf{x}) = -\frac{1}{c}\mathbf{x}^T(\mathbf{e}_k + \mathbf{e}_l) = -\frac{1}{c}\mathbf{x}^T\mathbf{e}_{kl}, \quad (1)$$

where $\mathbf{e}_k = [\cos\phi_k, \sin\phi_k]^T$ and $\mathbf{e}_l = [\cos\phi_l, \sin\phi_l]^T$ are unit vectors in the direction of the $k$-th transmitter and $l$-th receiver respectively. The signal delay to the location $\mathbf{x}$ in the direction of the $k$-th transmitter is $\tau_k(\mathbf{x}) = -\frac{1}{c}\mathbf{x}^T\mathbf{e}_k$ and similarly, the signal delay to the location $\mathbf{x}$ in the direction of the $l$-th receiver is $\tau_l(\mathbf{x}) = -\frac{1}{c}\mathbf{x}^T\mathbf{e}_l$. Thus, the propagation delay is determined by the projection of a scatterer's location onto the $kl$-th transmit-receive pair's bi-static range vector $\mathbf{e}_{kl} \doteq \mathbf{e}_k + \mathbf{e}_l$.

For extended scenes, multiple scatterers will have the same projection onto the bistatic range direction. The collection of such scatterers satisfies $\{\mathbf{x}|\frac{1}{2}\mathbf{x}^T\mathbf{e}_{kl} = \rho\}$. These scatterers are simultaneously illuminated and have their collective response $q_{kl}(\rho)$ registered at the receive antenna with the same delay. The so-called range profile $q_{kl}(\rho)$ is an aggregate response at each delay or range and is given by:

$$q_{kl}(\rho) = \int_{\|\mathbf{x}\| \le L} s(\mathbf{x})\delta\left(\rho - \frac{1}{2}\mathbf{x}^T\mathbf{e}_{kl}\right)d\mathbf{x}.$$

Under the far field and narrow-band transmit signal assumption [13], the overall received signal from the entire ground patch is assumed to be a superposition of the returns from all the scattering centers and is given by:

$$r_{kl}(t) = \int_{-L}^{L} q_{kl}(\rho)\gamma_k\left(t - \tau_{kl}(\mathbf{x}_o) + \frac{2\rho}{c}\right)d\rho. \quad (2)$$

In terms of the spatial reflectivity function $s(\mathbf{x})$ the received signal is given by:

$$r_{kl}(t) = \int_{\|\mathbf{x}\| \le L} s(\mathbf{x})\gamma_k\left(t - \tau_{kl}(\mathbf{x}_o) - \tau_{kl}(\mathbf{x})\right)d\mathbf{x}, \quad (3)$$

where $\tau_{kl}(\mathbf{x})$ is given in (1).

### B. Moving scene model

When moving scatterers are present in the scene we are interested in producing a focused image of the spatial reflectivity function at some reference time $t_{ref}$. We assume that the scatterer at location $\mathbf{x}$ has an associated arbitrary constant velocity vector $\mathbf{v}_{\mathbf{x}} = [v_x, v_y]^T$. Note that in the case of non-constant motion, the true scatter motion can be well approximated as constant when the time scale of the coherent processing interval (CPI) is relatively small. We consider the general case when the CPI interval contains multiple transmitted pulses.

Let us first consider the effect of the scatterer motion during one pulse transmission following the analysis developed in [13]. In the case of motion, the delay of the transmitted waveform is dependent on both the location and the velocity of the scatterer. The signal reflected from a single moving scatterer will now be of the form

$$r_{kl}(t) = s(\mathbf{x})\gamma_k\left(t - \tau_{kl}(\mathbf{x}_o, \mathbf{x}, \mathbf{v}_{\mathbf{x}})\right),$$

where $\tau_{kl}(\mathbf{x}_o, \mathbf{x}, \mathbf{v}_{\mathbf{x}})$ is the delay including the effects of motion. Next, we derive an expression for $t - \tau_{kl}(\mathbf{x}_o, \mathbf{x}, \mathbf{v}_{\mathbf{x}})$ in the case of motion.

Let $\hat{t}_{\mathbf{x}, \mathbf{v}_{\mathbf{x}}}(t)$ denote the time when the transmitted wave reaching the $l$-th receiver at time $t$ interacted with the scatterer, that at the reference time $\tau_k(\mathbf{x}_o)$ was located at $\mathbf{x}$. Recall that $\tau_k(\mathbf{x}_o)$ is the propagation delay between the $k$-th transmitter and the scene's origin. Thus, the scatterer at time $\tilde{t}_{\mathbf{x},\mathbf{v}_\mathbf{x}}(t)$ is actually located at $\mathbf{x} + \mathbf{v}_\mathbf{x}\tilde{t}_{\mathbf{x},\mathbf{v}_\mathbf{x}}(t)$. We can write

$$\tilde{t}_{\mathbf{x},\mathbf{v}_\mathbf{x}}(t) = t - \left[ \tau_l(\mathbf{x}_o) - \frac{\left[\mathbf{x} + \mathbf{v}_\mathbf{x}\tilde{t}_{\mathbf{x},\mathbf{v}_\mathbf{x}}(t)\right]^T \mathbf{e}_l}{c} \right]. \quad (4)$$

Solving the above equation for $\tilde{t}_{\mathbf{x},\mathbf{v}_\mathbf{x}}(t)$ and noting that $\tau_k(\mathbf{x}; \tilde{t}_{\mathbf{x},\mathbf{v}_\mathbf{x}}(t)) = -\frac{1}{c}\left[\mathbf{x} + \mathbf{v}_\mathbf{x}\tilde{t}_{\mathbf{x},\mathbf{v}_\mathbf{x}}(t)\right]^T \mathbf{e}_k$ and $\tau_l(\mathbf{x}; \tilde{t}_{\mathbf{x},\mathbf{v}_\mathbf{x}}(t)) = -\frac{1}{c}\left[\mathbf{x} + \mathbf{v}_\mathbf{x}\tilde{t}_{\mathbf{x},\mathbf{v}_\mathbf{x}}(t)\right]^T \mathbf{e}_l$, the argument of the delayed pulse is derived to be:

$$\begin{aligned}
t &- \tau_{kl}(\mathbf{x}_o, \mathbf{x}, \mathbf{v}_\mathbf{x}) = \\
&= t - \left[\tau_k(\mathbf{x}_o) + \tau_k(\mathbf{x}; \tilde{t}_{\mathbf{x},\mathbf{v}_\mathbf{x}}(t))\right] \\
&\quad - \left[\tau_l(\mathbf{x}_o) + \tau_l(\mathbf{x}; \tilde{t}_{\mathbf{x},\mathbf{v}_\mathbf{x}}(t))\right] \\
&= -\tau_k(\mathbf{x}_o) + \mathbf{x}^T \mathbf{e}_k/c \\
&\quad + \frac{1 + \mathbf{v}_\mathbf{x}^T \mathbf{e}_k/c}{1 - \mathbf{v}_\mathbf{x}^T \mathbf{e}_l/c}\left(t - \tau_l(\mathbf{x}_o) + \mathbf{x}^T \mathbf{e}_l/c\right) \\
&\approx t - \tau_{kl}(\mathbf{x}_o) - \tau_{kl}(\mathbf{x}) \\
&\quad + \frac{\mathbf{v}_\mathbf{x}^T \mathbf{e}_{kl}}{c}\left(t - \tau_k(\mathbf{x}_o) + \frac{\mathbf{x}^T \mathbf{e}_l}{c}\right),
\end{aligned}$$

where we used the fact that $|\mathbf{v}_\mathbf{x}|/c \ll 1$, such that $\frac{1 + \mathbf{v}_\mathbf{x}^T \mathbf{e}_k/c}{1 - \mathbf{v}_\mathbf{x}^T \mathbf{e}_l/c} \approx 1 + \mathbf{v}_\mathbf{x}^T \mathbf{e}_{kl}/c$.

Extending the model to the case of multiple probe transmissions during the coherent processing interval (CPI), we assume that the scene is imaged at some arbitrary time $t_{ref}$ outside the given pulse interval. Then, the scatterer located at $\mathbf{x}$ at the reference time $t_{ref}$, will be located at $\mathbf{x} + \mathbf{v}_\mathbf{x}(t_k - t_{ref})$ at time $t_k$ that represents the time when the $k$-th transmitted pulse reached the scene origin. Updating $\mathbf{x}$ of the previous equation with $\mathbf{x} + (t_k - t_{ref})\mathbf{v}_\mathbf{x}$, we obtain for the $k$-th transmitter pulse:

$$\begin{aligned}
t &- \tau_{kl}(\mathbf{x}_o, \mathbf{x}, \mathbf{v}_\mathbf{x}) \approx \\
&\quad t - \tau_{kl}(\mathbf{x}_o) - \tau_{kl}(\mathbf{x}) \\
&\quad + \frac{\mathbf{v}_\mathbf{x}^T \mathbf{e}_{kl}}{c}\left(t + t_k - t_{ref} + \epsilon_{kl}(\mathbf{x}, \mathbf{v}_\mathbf{x})\right),
\end{aligned} \quad (5)$$

where $\epsilon_{kl}(\mathbf{x}, \mathbf{v}_\mathbf{x}) = -\tau_k(\mathbf{x}_o) + [\mathbf{x} + (t_k - t_{ref})\mathbf{v}_\mathbf{x}]^T \mathbf{e}_l/c$.

Finally, the forward observation model in the presence of motion becomes:

$$r_{kl}(t) = \int_{\|\mathbf{x}\| \le L} s(\mathbf{x})\gamma_k\left(t - \tau_{kl}(\mathbf{x}_o, \mathbf{x}, \mathbf{v}_\mathbf{x})\right) d\mathbf{x}, \quad (6)$$

with the argument $t - \tau_{kl}(\mathbf{x}_o, \mathbf{x}, \mathbf{v}_\mathbf{x})$ given in (6) and $s(\mathbf{x})$ representing the spatial reflectivity function at the reference time. Comparing this equation to the received signal model for the stationary scene given by (3), we see that the two models differ by an additional delay attributed to the scatterers' motion. The additional delay of a scatterer present at the location $\mathbf{x}$ at the reference time $t_{ref}$ is proportional to the projection of the scatterers velocity $\mathbf{v}_\mathbf{x}$ to the $kl$-th transmit-receive pair bistatic range vector $\mathbf{e}_{kl}$ and the time interval in between the observation time and the reference time.

For narrowband waveforms, defined by $\gamma_k(t) = \tilde{\gamma}_k(t)e^{-j\omega_k t}$, where $\tilde{\gamma}_k(t)$ is the low-pass equivalent signal and $\omega_k$ the carrier frequency, (6) is approximated [13] by:

$$\begin{aligned}
r_{kl}(t) &\approx e^{-j\omega_k(t - \tau_{kl}(\mathbf{x}_o))} \\
&\int_{\|\mathbf{x}\| \le L} s(\mathbf{x})\tilde{\gamma}_k\left(t - \tau_{kl}(\mathbf{x}_o) - \tau_{kl}(\mathbf{x})\right) \\
&\quad e^{-j\omega_k\left[-\tau_{kl}(\mathbf{x}) + [t + t_k - t_{ref} + \epsilon_{kl}(\mathbf{x}, \mathbf{v}_\mathbf{x})]\frac{\mathbf{v}_\mathbf{x}^T \mathbf{e}_{kl}}{c}\right]} d\mathbf{x}. \quad (7)
\end{aligned}$$

The additional phase shift of (7) is a function of the quantity $\omega_k \frac{\mathbf{v}_\mathbf{x}^T \mathbf{e}_{kl}}{c}$, which is basically the Doppler shift. Thus, the received signal is in the familiar form of the superposition of time-delayed and Doppler-shifted replicas. Notice that the Doppler shift is unique for each transmit-receive pair as it depends on the bistatic range vector $\mathbf{e}_{kl}$.

### C. Discrete model

A discrete version of the model in (6) or (7) can be obtained by discretizing the spatial variable $\mathbf{x}$ and sampling in time which, in the presence of receiver noise $\mathbf{n}$, becomes:

$$\mathbf{r} = \sum_{p=1}^{P} \mathbf{\Phi}_p(\mathbf{v}_p)s_p + \mathbf{n} = \mathbf{\Phi}(\mathbf{V})\mathbf{s} + \mathbf{n}. \quad (8)$$

In this equation, $\mathbf{r}$ represents the observed, thus known, set of return signals at all receivers across time. Its elements are indexed by the tuple $(k, l, t_s)$, with $t_s$ being the sampling times associated with the $kl$-th transmit-receive pair. The reflectivity of the $p$-th spatial cell or pixel is denoted by $s_p \in \mathcal{C}^{1 \times 1}$ and $\mathbf{\Phi}_p(\mathbf{v}_p)$ is the vector capturing the contribution to the received signal of a reflector that was located in the $p$-th pixel at the reference time $t_{ref}$ and moved with the constant velocity $\mathbf{v}_p$ throughout the coherent processing interval. Stationary point reflectors are included in this model by simply setting $\mathbf{v}_p = \mathbf{0}$.

The received signal model described in (8), represents the observation model of unknown scatterers' reflectivity coefficients $s_p$ and their corresponding velocity vectors $\mathbf{v}_p$. While the scattering coefficients $s_p$ enter the problem linearly, the unknown velocities $\mathbf{v}_p$ do not, so the overall problem is nonlinear and coupled. When the velocities are known the remaining equation for $s_p$ is linear, however, and straight forward focusing and estimation of scattering coefficients is possible. When the velocity is ignored (set to zero) or set to an incorrect value, the resulting reconstruction exhibits defocusing of the energy of the moving scatterer [1], [4]. These observations lead to one conventional approach to solving this problem. In particular, reconstructions are performed for every possible velocity yielding a large space-velocity cube [9], [2], [4]. Velocity slices where the image is well focused are assumed to indicate the correct velocity at a pixel. This approach requires the decoupled solution of many inversion problems and results in a large, somewhat ambiguous output. Recently, an approach based solely on the inversion of the forward operator is presented in [13], where a filtered-backprojection approach to imaging of moving targets takes the form of a weighted matched filtering which still produces a large space-velocity cube. In this paper we present a different approach based on recent results in sparse signal representation using overcomplete dictionaries [15], sparsity based reconstruction, and compressed sensing [16], [17], which is described next.

## III. Overcomplete dictionary approach

Sparse signal representation aims at capturing a complicated signal as a linear combination of a few generating elements [15]. In particular, a signal in a given class of dimension $M$ should be representable by a small subset of a collection of $Q$ generating elements. For the case when $M = Q$ and the elements are independent, the collection is termed a basis. When $Q > M$ the collection is termed an overcomplete basis or dictionary. To represent this problem mathematically, let $\mathbf{r} \in \mathcal{R}^{M \times 1}$ represent the signal, $\boldsymbol{\Phi} \in \mathcal{R}^{\mathbf{M} \times \mathbf{Q}}$ the dictionary and $\mathbf{s} \in \mathcal{R}^{Q \times 1}$ represent the linear coefficients, such that $\mathbf{r} = \boldsymbol{\Phi}\mathbf{s}$. Since $\boldsymbol{\Phi}$ has a non-zero null space many solutions are possible. What is sought is a sparse solution with only a few non-zero elements.

While optimal design of dictionaries $\boldsymbol{\Phi}$ is a topic of general interest, in this work we assume the dictionary is given and fixed based on prior knowledge of the expected velocities in a scene. The problem is then one of finding an optimally sparse solution. A direct formulation of this problem can be given as:

$$\min_{\mathbf{s}} \|\mathbf{s}\|_0 \qquad \text{s.t.} \quad \mathbf{r} = \boldsymbol{\Phi}\mathbf{s},$$

where $\|\cdot\|_0$ denotes the $l_0$ norm, which counts the number of non-zero elements of the argument. Unfortunately, this formulation is computationally difficult to solve, as it involves NP-hard enumerative search which is prohibitively expensive for even moderate sizes of $Q$. A number of alternative, indirect techniques have been developed, based either on relaxation techniques or iterative greedy algorithms.

The convex relaxation approach relies on the fact that besides the $l_0$ norm, the $l_1$ norm also promotes sparsity in a solution. The $l_1$ norm is defined as $\|\mathbf{s}\|_1 = \left( \sum_{i=1}^{Q} |(\mathbf{s})_i| \right)$, where $(\mathbf{s})_i$ is the $i$-th element of $\mathbf{s}$. This norm is a convex function of its arguments. The relaxed version of the problem then takes the form:

$$\min_{\mathbf{s}} \|\mathbf{s}\|_1 \qquad \text{s.t.} \quad \mathbf{r} = \boldsymbol{\Phi}\mathbf{s},$$

which is essentially a linear program (LP). The use of this formulation has also been motivated by the fact that under certain conditions on the overcomplete basis $\boldsymbol{\Phi}$, the original problem and the relaxed version can be shown to have the same solution [18].

When the signal $\mathbf{r}$ is noisy, the signal representation problem becomes a signal approximation problem. The convex relaxation formulation of the noisy signal approximation problem is given by:

$$\min_{\mathbf{s}} \|\mathbf{s}\|_1 \qquad \text{s.t} \quad \|\mathbf{r} - \boldsymbol{\Phi}\mathbf{s}\|_2^2 \leq \delta$$

where $\delta$ represents a small noise allowance. Instead of satisfying the relationship exactly, the solution coefficient vector $\mathbf{s}$ is allowed to satisfy the relationship approximately. This problem is known in the literature as noisy basis pursuit [19]. We make use of this formulation in what follows.

Compressed sensing (CS) [16], [17], takes the sparse representation framework one step further by seeking to acquire as few measurements as possible about a sparse signal, and given these measurements, reconstruct the sparse signal either exactly or with provably small probability of error using formulations like the above. Most of the work in CS assumes that the projections are drawn at random. However it is also known that Fourier measurements represent good projections for compressed sensing of sparse point like signals [17]. This result immediately connects to radar measurements, as SAR sensors can be viewed as measuring samples of the stationary scattering field in the spatial frequency domain.

### A. The new formulation

We exploit the overcomplete dictionary approach to signal representation to create a new formulation of the dynamic SAR inversion problem. In particular, we first introduce an appropriate overcomplete dictionary of velocity hypotheses by constructing a grid of all possible scatterer velocities at each location. This defines an over-complete representation of the received signal $\mathbf{r}$. This representation is combined with the SAR observation equation resulting in an inverse problem that is linear in terms of an extended reflectivity coefficient vector, but is now under-determined. We then apply a modified version of the sparsity seeking formulation of (11) to solve the resulting large, under-determined linear problem. Details are given next.

*Step 1: Dictionary Definition.* We hypothesize that a scatterer velocity $\mathbf{v}_p, \forall p$ belongs to one of a discrete set of velocities $\tilde{\mathbf{V}}$:

$$\mathbf{v}_p \in \tilde{\mathbf{V}} = \{\tilde{\mathbf{v}}_1 = 0, \tilde{\mathbf{v}}_2, \tilde{\mathbf{v}}_3, \ldots, \tilde{\mathbf{v}}_N\},$$

where $N$ denotes the size of the velocity grid. The first velocity vector is set to zero to allow for stationary targets. The original observation vector $\boldsymbol{\Phi}_p(\mathbf{v}_p)$ describing the contribution of the $p$-th pixel to the received signal $\mathbf{r}$ in terms of an *unknown* pixel velocity $\mathbf{v}_p$ now becomes the matrix of $[\boldsymbol{\Phi}_p(\tilde{\mathbf{v}}_1), \boldsymbol{\Phi}_p(\tilde{\mathbf{v}}_2), \ldots \boldsymbol{\Phi}_p(\tilde{\mathbf{v}}_N)]$ composed of the contribution of each possible *known* velocity hypothesis at pixel $p$ to the received signal. There are no unknowns in this matrix, in contrast to the original observation vector.

Finally, by combining this model at each pixel we obtain an overall overcomplete forward operator:

$$\boldsymbol{\Phi}_{\tilde{\mathbf{V}}} = [\boldsymbol{\Phi}_1(\tilde{\mathbf{v}}_1), \ldots, \boldsymbol{\Phi}_1(\tilde{\mathbf{v}}_N), \ldots, \boldsymbol{\Phi}_P(\tilde{\mathbf{v}}_1), \ldots, \boldsymbol{\Phi}_P(\tilde{\mathbf{v}}_N)].$$

To match this change, the reflectivity coefficient at pixel $p$ is accordingly expanded to become a reflectivity coefficient vector:

$$s_p \rightarrow \mathbf{s}_p^{\mathbf{b}} = s_p \begin{bmatrix} b_{p1} \\ b_{p2} \\ \vdots \\ b_{pN} \end{bmatrix},$$

where the auxiliary variables $b_{pn}$ are constrained to the binary set $b_{pn} \in \{0, 1\}$ to represent a 'true' or 'false' hypothesis that the $p$-th spatial location moves with $n$-th quantized velocity vector $\tilde{\mathbf{v}}_n$. Additionally, the variables $b_{pn}$ need to satisfy $\sum_n b_{pn} = 1$ for each $p$ to recover the model of (8). This additional constraint specifies that at most one spatial reflector is present within the $p$-th resolution cell at the reference time. Other, more complicated, models are also possible.

Stacking up these single-pixel reflectivity coefficients yields an overall, extended reflectivity vector for the entire image:

$$\mathbf{s_b} = \begin{bmatrix} \mathbf{s_1^b} \\ \mathbf{s_2^b} \\ \vdots \\ \mathbf{s_P^b} \end{bmatrix}.$$

Combining this extended reflectivity vector with the overcomplete forward operator yields our new overall linear motion-SAR forward model:

$$\mathbf{r} = \mathbf{\Phi_{\widetilde{V}}} \mathbf{s_b} + \mathbf{n} + \mathbf{q}, \tag{9}$$

where $\mathbf{s_b}$ is the extended reflectivity coefficient and $\mathbf{q}$ represents an additional noise term due to the velocity space quantization. In this new model $\mathbf{\Phi_{\widetilde{V}}}$ is completely specified and contains no unknowns. We have essentially converted the original, difficult, non-linear problem to a selection problem with a linear observation.

*Step 2: Image Formation.* We now treat the problem of image formation as a problem of inverting (9). In principle, we seek to find a sparse vector $\mathbf{s_b}$ that best describes the received signal. In particular, we seek a solution of the following optimization problem:

$$\begin{aligned} \min_{\mathbf{s_b}, \mathbf{b}, \mathbf{s}} \quad & \|\|\mathbf{s_b}\|\|_0 \\ \text{subject to} \quad & \|\mathbf{r} - \mathbf{\Phi_{\widetilde{V}}} \mathbf{s_b}\|_2^2 \leq \delta \\ & (\mathbf{s_b})_{(p-1)N+n} = s_p \cdot b_{pn} \\ & \sum_n b_{pn} = 1 \qquad p = 1, \ldots, P \\ & b_{pn} \in \{0, 1\} \qquad n = 1, \ldots N. \end{aligned}$$

where $\delta$ represents some noise allowance aimed at capturing both receiver and quantization noise and the $l_0$ norm $\|\cdot\|_0$ counts the number of nonzero components in its argument. Since the coherence in typical SAR scenes is contained in the field magnitude (due to random phase), the $l_0$ norm is applied explicitly to magnitudes of the extended complex reflectivity field $|\mathbf{s_b}|$.

### B. Solving the formulation

The above optimization problem is a non-convex, mixed-integer program, known to be NP hard to solve. Thus, we resort to a two step procedure based on convex relaxation for its approximate solution.

Step 1: We first solve the following relaxed optimization problem:

$$\begin{aligned} \widehat{\mathbf{s}}_\mathbf{b} \quad = \quad & \arg\min_{\mathbf{s_b}} \qquad \|\|\mathbf{s_b}\|\|_1 \\ & \text{subject to} \quad \|\mathbf{r} - \mathbf{\Phi_{\widetilde{V}}} \mathbf{s_b}\|_2^2 \leq \epsilon, \end{aligned}$$

where $\|\|\mathbf{s_b}\|\|_1 = \sum_{i=1}^{PN} \sqrt{(\Re(\mathbf{s_b})_i)^2 + (\Im(\mathbf{s_b})_i)^2}$ and $(\mathbf{s_b})_i$ is the $i$-th element of the vector $\mathbf{s_b}$. Note in this step we use the convex $l_1$ penalty and do not enforce the exclusivity of velocity hypotheses represented by the auxiliary binary variables $b_{pn}$.

Step 2: We then find the final reflectivity and velocity vector estimates by selecting the maximum magnitude response at the $p$-th pixel:

$$\begin{aligned} (\widehat{s}_p, \widehat{n}) \quad &= \quad \arg\max_n |(\widehat{\mathbf{s}}_\mathbf{b})_{(p-1)N+n}|, \\ \widehat{\mathbf{v}}_p \quad &= \quad \widetilde{\mathbf{v}}_{\widehat{n}}. \end{aligned} \tag{10}$$

This is consistent with the assumption that only one moving reflector is present in the $p$-th pixel at the reference time.

By constructing the overcomplete dictionary and applying convex relaxation we have linearized the forward model at the expense of an increase in the size of the problem. The minimization problem in Step 1 is a second-order cone program, which we solve by a specialized large scale interior-point method for complex variables proposed first in [20]. The method is a specialized central path interior-point method with an approximate search direction found through a pre-conditioned conjugate gradient method which yields efficient solution of such large problems.

Our approach relies on two levels of sparsity. First, the introduction and use of an overcomplete dictionary mandates that we seek a sparse solution vector $\mathbf{s_b}$. In addition, compressed sensing theory relates the number of measurements necessary for accurate recovery of $\mathbf{s_b}$ to its underlying sparsity [16], [17], indicating that it may be possible to perform accurate recovery of sparse scenes with relatively few transmitters and receivers.

We also want to emphasize that as long as the scene contains a sparse set of reflectors, there is no constraint on the object velocity the method can handle. The velocity grid resolution should simply be matched to the coherent processing interval and the maximal carrier frequency of the transmitted signal in order to avoid phase wrapping of the exponent in (7). The phase error due to the velocity quantization is linear in $\mathbf{v_x}$, so the velocity grid need not be constant: at smaller velocity magnitudes it can be coarser and at higher velocity magnitudes it can be finer. Finally, the discrete model of (8) implies that only a single scatterer is present at the reference time in each spatial pixel. For the case when the spatial grid is sufficiently coarse to include multiple point reflectors within the resolution cell, (8) can be written as:

$$\mathbf{r} = \sum_{m=1}^{M(p)} \sum_{p=1}^{P} \mathbf{\Phi}_p(\mathbf{v}_{p,m}) s_{p,m} + \mathbf{n}, \tag{11}$$

where $\mathbf{v}_{p,m}$ models the velocity of the $m$-th reflector in the $p$-th resolution cell, that contains $M(p)$ scatterers out of which one is allowed to be stationary. The algorithm accommodates the new model with a small change. In case of $M(p)$ reflectors per spatial grid cell, the constraint $\sum_n b_{pn} = 1$, should be replaced with $\sum_n b_{pn} = M(p)$. After convex relaxation, (10) should be accordingly modified to retain $M(p)$ maximal magnitude reflectivity values and their corresponding velocity vectors.

### IV. TRANSMITTED WAVEFORMS

We presented the overcomplete dictionary reconstruction algorithm without specifying the transmitted signals. The formulation we presented is quite general and can potentially work for many waveforms and sensor configurations.

In this initial work, to demonstrate the algorithm we consider multiple snapshot, non-overlapping transmissions of chirp and ultra-narrow band signals. Such signals are known to lead to a simple Fourier relationship between the stationary reflectivity field and the measurements for both monostatic and multiple distributed antenna configurations [14], [9], [21]. The corresponding forward operators posses compressed sensing properties, allowing for good reconstruction of sparse fields with few measurements [22], [17]. In the following, we discuss the specific forward radar model resulting from transmission of such waveforms.

The chirp signal is the most common spotlight SAR pulse [14], given by

$$\gamma_k(t) = \begin{cases} e^{j\alpha_k t^2} \cdot e^{j\omega_k t}, & -\frac{\tau_c}{2} \leq t \leq \frac{\tau_c}{2} \\ 0 & \text{otherwise,} \end{cases}$$

where $\omega_k$ is the center frequency and $\alpha_k$ is the so-called chirp rate of the $k$-th transmit element. The frequencies encoded by the chirp signal extend from $\omega_k - \alpha_k \tau_c$ to $\omega_k + \alpha_k \tau_c$, such that the bandwidth of this signal is given by $B_k = \frac{\alpha_k \tau_c}{\pi}$. Ultra-narrow band waveforms are special cases of the chirp signal obtained by setting $\alpha_k = 0$.

We use this transmitted chirp signal in (6) and apply typical demodulation and baseband processing. In particular, the received signal is mixed with the transmitted signal referenced to the origin of the scene $e^{-j[\omega_k(t - \tau_{kl}(\mathbf{x}_o)) + \alpha_k(t - \tau_{kl}(\mathbf{x}_o))^2]}$, and then low-pass filtered. If the quadratic phase error is ignored [14], we obtain the following signal as the input to our algorithm:

$$r_{kl}(t) \approx \int_{\|\mathbf{x}\| < L} s(\mathbf{x}) \, e^{-j\Omega_{kl}(t)\mathbf{x}^T \mathbf{e}_{kl}}$$
$$e^{-j\Omega_{kl}(t)[t + t_k - t_{ref} + \epsilon_{kl}(\mathbf{x}, \mathbf{v}_\mathbf{x})]\mathbf{v}_\mathbf{x}^T \mathbf{e}_{kl}} \, d\mathbf{x}, \quad (12)$$

where $\Omega_{kl}(t) = \frac{1}{c}[\omega_k - 2\alpha_k(t - \tau_{kl}(\mathbf{x}_o))]$, depends on the frequency content of the transmitted waveform.

The first exponential term depends on the stationary scatters only. The second exponential term depends on the moving scatters only. For stationary scenes with $\mathbf{v}_\mathbf{x} = \mathbf{0}, \forall \mathbf{x}$, (12) represents the 2D Fourier transform of the spatial reflectivity function, evaluated at the discrete set of the spatial frequency vectors $\mathbf{k} \doteq [k_x, k_y]^T$ given by

$$\mathbf{k}_{k,l,t} = \Omega_{kl}(t)\mathbf{e}_{kl} = \Omega_{kl}(t)(\mathbf{e}_k + \mathbf{e}_l). \quad (13)$$

This equation can be used to describe the spatial frequency sampling of both the monostatic and the multi-static configuration [9]. Recall that $\mathbf{e}_k$ and $\mathbf{e}_l$ are the unit vectors in the direction of the $k$-th transmitter and the $l$-th receiver. For the monostatic case these two vectors coincide, i.e. $\mathbf{e}_k = \mathbf{e}_l$. For the monostatic case and fixed $t$, (13) represents an arc of the circle of the radius $\Omega_{kl}(t)$ centered the origin of the spatial frequency domain, with the length of the arc determined by the radial span of the vector $\mathbf{e}_k$. Changing $t$, expands or shrinks the circular arcs, leading to the familiar key-hole sampling [14] of the conventional monostatic case.

For the multi-static case, the situation is a bit more complicated. For simplicity, assume that $\Omega_{kl}(t) = \Omega(t)$. At fixed $t$ and fixed $\mathbf{e}_k$, (13) also describes an arc of a circle, this time passing through the origin and centered at $\Omega(t)$ along the direction $\mathbf{e}_k$. The length of the arc and its orientation are determined by the radial span of the receiver vector $\mathbf{e}_l$. Changing $t$ results in expansion or shrinkage of the circle passing through the origin with its center sliding along the direction $\mathbf{e}_k$.

Thus, different covering patters of $\mathbf{k}$-space are possible with various sampling strategies in space and time, while the resolution is primarily determined by the extent of the $\mathbf{k}$-space covering. The range and cross-range resolution of both the conventional monostatic SAR system and the multi-static, distributed antenna SAR with antenna elements confined within the forward cone of $\Delta \theta < \pi/2$ are lower bounded by the bounding box of the annulus:

$$\rho_x \geq \frac{c}{2B_{eq}}, \quad \rho_y \geq \frac{c}{4(f_o + B/2)\sin(\Delta\theta/2)}, \quad (14)$$

where $B_{eq} = (f_0 + B/2) - (f_0 - B/2)\cos(\Delta\theta/2)$. We illustrate the $\mathbf{k}$-space sampling patterns used in our experiments in Fig. 2 and Fig. 3. Fig. 2(a) and Fig. 2(b) show the $\mathbf{k}$-space sampling of the conventional monostatic SAR with bandwidth of $B = 50$MHz, centered at $f_0 = 1.5$GHz, over a narrow synthetic aperture of $\Delta\theta = 5 \deg$ and a wide synthetic aperture of $\Delta\theta = 45 \deg$, respectively. The $\mathbf{k}$-space covering of the narrow-angle monostatic SAR is well approximated with the circumscribed rectangle, while this is not the case for the wide-angle monostatic SAR.

A $\mathbf{k}$-space covering similar to the wide-angle mono-static case in Fig. 2(b) can be achieved with a multi-static, distributed transmit and receive antenna using continuous wave transmission. Fig. 3 illustrates the $\mathbf{k}$-space covering for two multi-static distributed antenna configurations utilizing continuous wave transmission with transmitters and receivers positioned within the forward cone of $\Delta\theta = 45 \deg$. Fig. 3(a) shows a multi-static configuration where transmitters sequentially transmit a single tone of frequency $f_0 = 1.5$GHz. Fig. 3(b) shows a multi-static configuration where different transmitters send out different continuous wave signals within a bandwidth of $50$MHz around the center frequency $f_0 = 1.5$GHz. This sampling can be achieved either via sequential multi-static transmission or via simultaneous MIMO transmission, as such signals are easily separated at each receiver. Although the expected resolutions in the last three cases are similar, it is important to notice that the $\mathbf{k}$-space covering of Fig. 3(b) for MIMO transmission is achieved in the least amount of time, making it the most favorable configuration for imaging of moving objects.

## V. NUMERICAL EXPERIMENTS

In this section we outline several numerical experiments demonstrating the reconstruction capability of the overcomplete dictionary approach for imaging scenes that contain both stationary and moving objects. First, we show results for a multi-static, distributed antenna sensing configuration for cases corresponding to a stationary scene, a scene with moving objects whose motion is ignored, a scene with moving objects whose motion is explicitly handled by the overcomplete dictionary reconstruction algorithm. Next, we show reconstruction results for the same cases, but for a wide-angle monostatic configuration with the same lower bounds on the range and the cross range resolution. Finally, we compare our approach with the matched filtering approach described in [9] and [13].

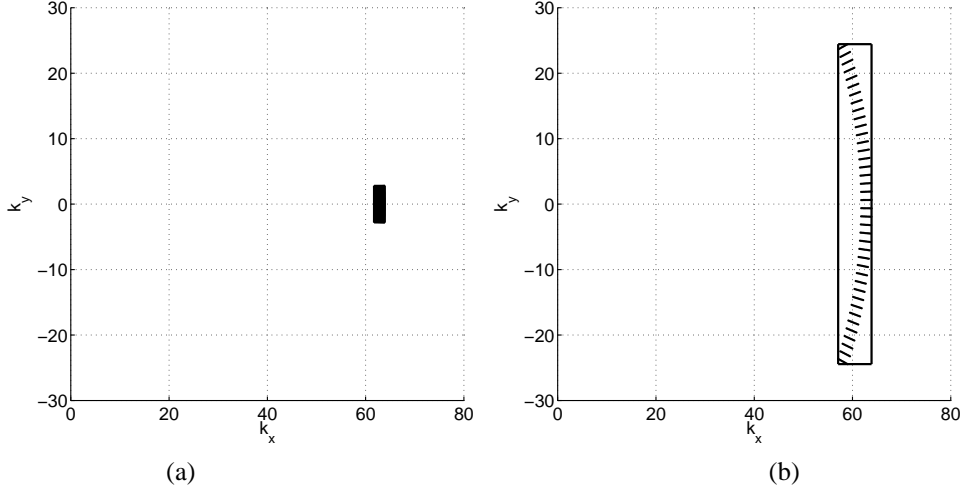

Fig. 2. Mono-static SAR **k**-space sampling for $B = 50$MHz, $f_0 = 1.5$GHz with the forward cone centered at 0 deg (a) Conventional, narrow-angle SAR with $\Delta\theta = 5$ deg, $(\rho_x \geq 2.9\text{m}, \rho_y \geq 1.13\text{m})$ and (b) Wide-angle SAR with $\Delta\theta = 45$ deg, $(\rho_x \geq 0.9\text{m}, \rho_y \geq 0.13\text{m})$ .

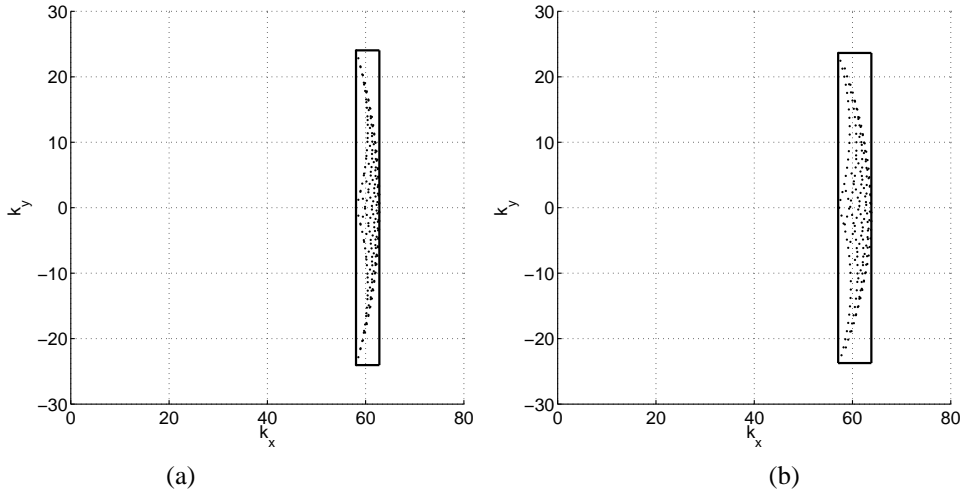

Fig. 3. Multi-static, distributed antenna SAR **k**-space sampling with receive and transmit antenna elements positioned within the forward cone of $\Delta\theta = 45$ deg centered at 0 deg. (a) The case where each transmitter uses continuous waves of frequency $f_0 = 1.5$GHz, $(\rho_x \geq 1.32\text{m}, \rho_y \geq 0.13\text{m})$ and (b) The case where each transmitter sends out continuous waves of different frequencies within the bandwidth of $B = 50$MHz centered at $f_0 = 1.5$GHz, $(\rho_x \geq 0.9\text{m}, \rho_y \geq 0.13\text{m})$ .

For the multi-static, distributed antenna configuration, all transmit and receive elements of the synthetic aperture are positioned in the forward cone of $\Delta\theta = 45$ deg with its center direction aligned with the $x$-axis of the coordinate system, as illustrated in Fig. 1. We choose a relatively narrow forward cone in order to better accommodate the isotropic scattering assumption of realistic scatterers. Each antenna element transmits a distinct continuous wave signal with a frequency randomly chosen within the bandwidth of $B = 50$MHz around the center frequency of $f_0 = 1.5$GHz. Similarly, for the monostatic case, the angular extent of the synthetic aperture is also $\Delta\theta = 45$ deg with the center aspect aligned with the $x$-axis. The transmit waveform is chosen to be the conventional chirp signal of $B = 50$MHz centered at $f_0 = 1.5$GHz. The resulting **k**-space sampling patterns are shown in Fig. 2(b) and Fig. 3(b) for the monostatic and the multi-static distributed antenna configurations, respectively.

The scene of interest is of $32 \times 32$m in size, represented by $32 \times 128$ pixels in the $x$ and $y$ direction respectively, such that the spatial cell size is $(\Delta x, \Delta y) = (1, 0.25)$m. The scene contains two rigid objects in motion and one rigid stationary object, as illustrated in Fig. 4, which displays the ground truth images. An object is defined as a set of clustered pixels reflecting electromagnetic energy isotropically within the cone of $45$ deg. There are 20 active scatterers in each object. The magnitudes of velocities of the moving objects are 32.5m/s and 4.7m/s in the direction of $\pi/6$ and $\pi/6 + \pi/2$, respectively.

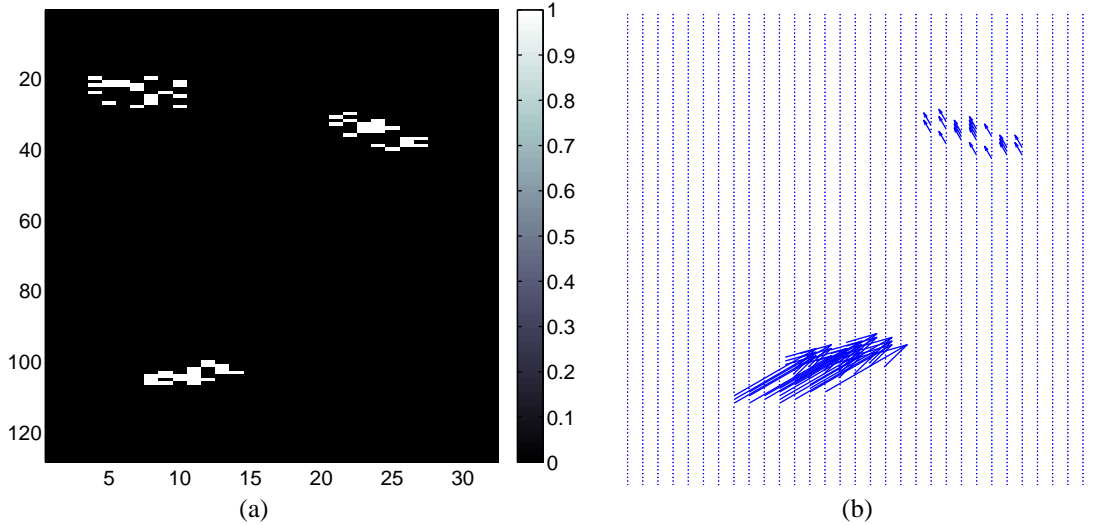

Fig. 4. The ground truth of the scene containing one stationary object (upper left corner) and two rigid objects moving at 32.5m/s (lower left corner) and 4.7m/s (upper right corner): (a) The reflectivity magnitude and (b) The velocity vectors associated with different pixels.

The reflectivity magnitude of the point scatterers in the scene is shown in Fig. 4(a) and their corresponding velocity vectors are shown in Fig. 4(b). All measurements are corrupted by independent Gaussian receiver noise such that $SNR = 20$dB, with $SNR$ defined as $SNR = 20 \log \frac{\|\Phi(V)x\|_2}{\|y - \Phi(V)x\|_2}$.

First, we show a set of reconstruction results for the multi-static, distributed antenna configuration with $N_{rx} = 40$ receive elements and $N_{tx} = 10$ transmit elements, each sequentially transmitting a distinct tone signal with a pulse repetition interval of $PRI = 2$ms. We start by considering what happens when there is no motion and when motion is ignored. In Fig. 5(a) the scene of interest is made completely stationary and our reconstruction is performed with the velocity dictionary $\tilde{V}$ containing only the velocity vector $v_1 = 0$, corresponding to a static scene. In this case, we see that all objects are well focused in the reconstruction as would be expected. These sparse multi-static stationary scene results essentially extend those in presented in [23] for the case of the conventional narrow-angle monostatic SAR. In Fig. 5(b) the scene is now made dynamic, as described by Fig. 4, but scatterers' velocities are ignored in the reconstruction, i.e. the dictionary $\tilde{V}$ again contains only the vector $v_1 = 0$. We see that when the motion of the objects is ignored, the stationary object still achieves reasonable focus, while the moving objects appear severely blurred.

Next we demonstrate what happens when we use our over-complete dictionary approach to capture the object velocities in dynamic scenes. In Fig. 5(c) we show the reflectivity magnitude reconstruction when the velocity dictionary contains velocities with a magnitude resolution of 3m/s in the range $[0, 10]$m/s and a resolution of 1m/s in the range $[30, 40]$m/s. The true object velocities are not part of this dictionary. The whole dynamic scene is re-focused with scatterer locations correctly identified, as illustrated in Fig. 5(d). All targets appear focused and accurately localized. In Fig. 6 we show the corresponding estimated target velocities. All reconstructed velocities are correctly estimated within the resolution grid error. The coarseness of the velocity grid is chosen to avoid phase wrapping and further, the phase deviation is minimized for shorter CPIs which, in return, are easier to support with multi-static and MIMO configurations.

We now repeat these experiments for the wide-angle mono-static configuration. Results are presented in Fig. 7 and Fig. 8, showing reconstructed magnitudes and velocities, respectively. Recall that each transmitted pulse is now a chirp signal with $B = 50$MHz at $f_0 = 1.5$GHz. The number of transmitted probes within the angular extent of $\Delta\theta = 45 \deg$ is 40, with each pulse return sampled at 10 frequencies. The pulse repetition interval is kept as before for consistency (we do not worry about the platform velocity required to transverse the angular extent within the coherent processing interval). The results presented for reconstruction of the dynamic scene are thus optimistic. We observe that these reconstructions are inferior to those obtained in the multi-static scenario, as some features of objects are blurred. The additional diversity provided by the multi-static configuration apparently translates into improved robustness and quality of the reconstruction.

Finally, we show reconstruction results for the weighted matched filtering approach described in [9] and [13] for the multi-static, distributed antenna configuration. The reflectivity magnitude is evaluated at each pixel for every hypothesized velocity, leading to a large space-velocity cube. Here we show the maximal value of the reflectivity magnitude of each pixel across the velocity direction in this space-velocity cube.

Fig. 9 shows the results obtained from matched filtering when $N_{rx} = 40$ receivers and $N_{tx} = 10$ transmitters are used with ultra-narrow band waveforms, as before. This configuration results in 400 observations. Fig. 9a) is the slice of the matched filter result at zero velocity, showing what happens when velocity is ignored. Fig. 9b) is the matched filter reconstruction of the scene ignoring motion. Fig. 9c) is the result obtained by taking the maximum response across the space-velocity cube and Fig. 9d) shows scatter localization, by thresholding the result of Fig. 9c) at 0.2 . For the same amount of data, the matched-filter-based reconstructions are much worse than those provided by our new overcomplete dictionary approach.

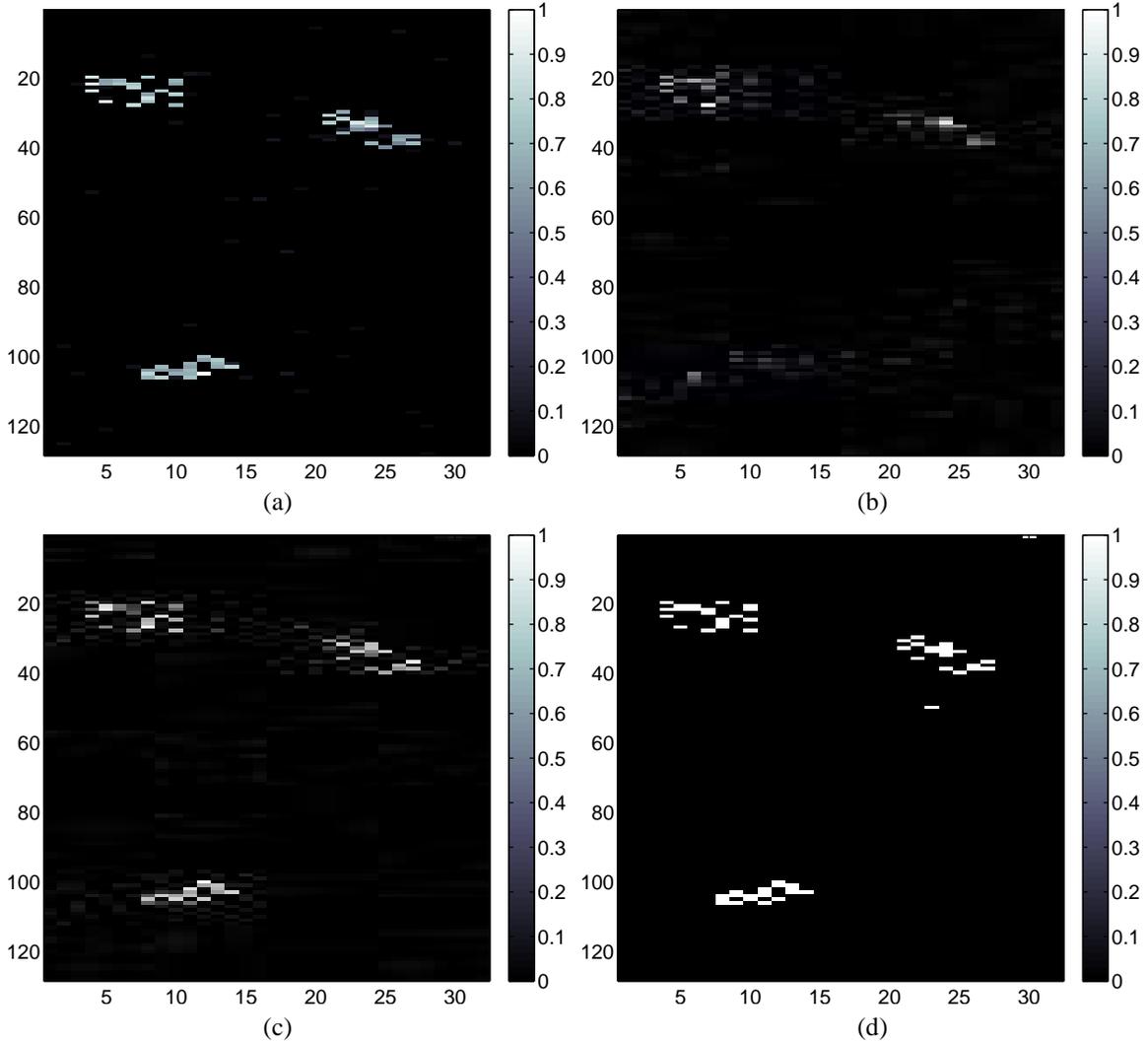

Fig. 5. Reflectivity magnitude reconstruction with our new, overcomplete dictionary approach for the multi-static, distributed antenna configuration at $SNR = 20$dB with 400 measurements, $(N_{tx}, N_{rx}) = (10, 40)$: (a) The reconstruction of the stationary scene assuming no motion. (b) The reconstruction of the dynamic scene when velocities are ignored, ($\tilde{\mathbf{V}} = \{\mathbf{v}_1 = \mathbf{0}\}$). (c) The reconstruction of the dynamic scene with the full overcomplete velocity dictionary. (d) The corresponding locations of reflectors in the reconstruction of (c) whose magnitudes are greater than 0.2.

To obtain matched-filter-based reconstructions that have similar quality to those produced by the overcomplete dictionary approach, we can increase the amount of data by using $N_{rx} = 40$ receivers and $N_{tx} = 10$ transmitters, but with the transmitters now emitting chirp waveforms of $B = 50$MHz at $f_0 = 1.5$GHz with $N_f = 30$ samples per waveform. Fig. 10 shows the match-filter-based results for this configuration, which corresponds to 12,000 observations. Fig. 10a) is the matched filter reconstruction of the static scene. Fig. 10b) is the matched filter reconstruction of the scene ignoring motion. Fig. 10c) is the maximum magnitude reflectively response of the matched filter reconstruction and Fig. 10d) is the thresholded version of this result, showing scatter localization. We see that the matched filter reconstruction can come close to recovering fine object features, but at the expense of significant increase in measurement data relative to the overcomplete dictionary approach.

In Table I we summarize these results by providing the per-pixel reflectivity magnitude error for the different cases, where this error is defined as $E = \|\mathbf{x} - \hat{\mathbf{x}}\|_2^2 / P$, with $\mathbf{x}$ the ground truth reflectivity magnitude, $\hat{\mathbf{x}}$ its estimate and $P$ is the number of pixels in the scene. The over-complete dictionary results are for 400 observations and the matched-filter results are for the 400 observation case and the expanded 12,000 observation case. The estimates generated by our new overcomplete dictionary method are the most accurate, producing over a several ten-fold and a 55% reduction in the error generated by the matched filter solution which uses the same amount of data and 30 times as much data, respectively. Further, the spatial diversity of the multi-static, distributed antenna configuration provides good reconstructions even in the case of narrowband transmission waveforms.

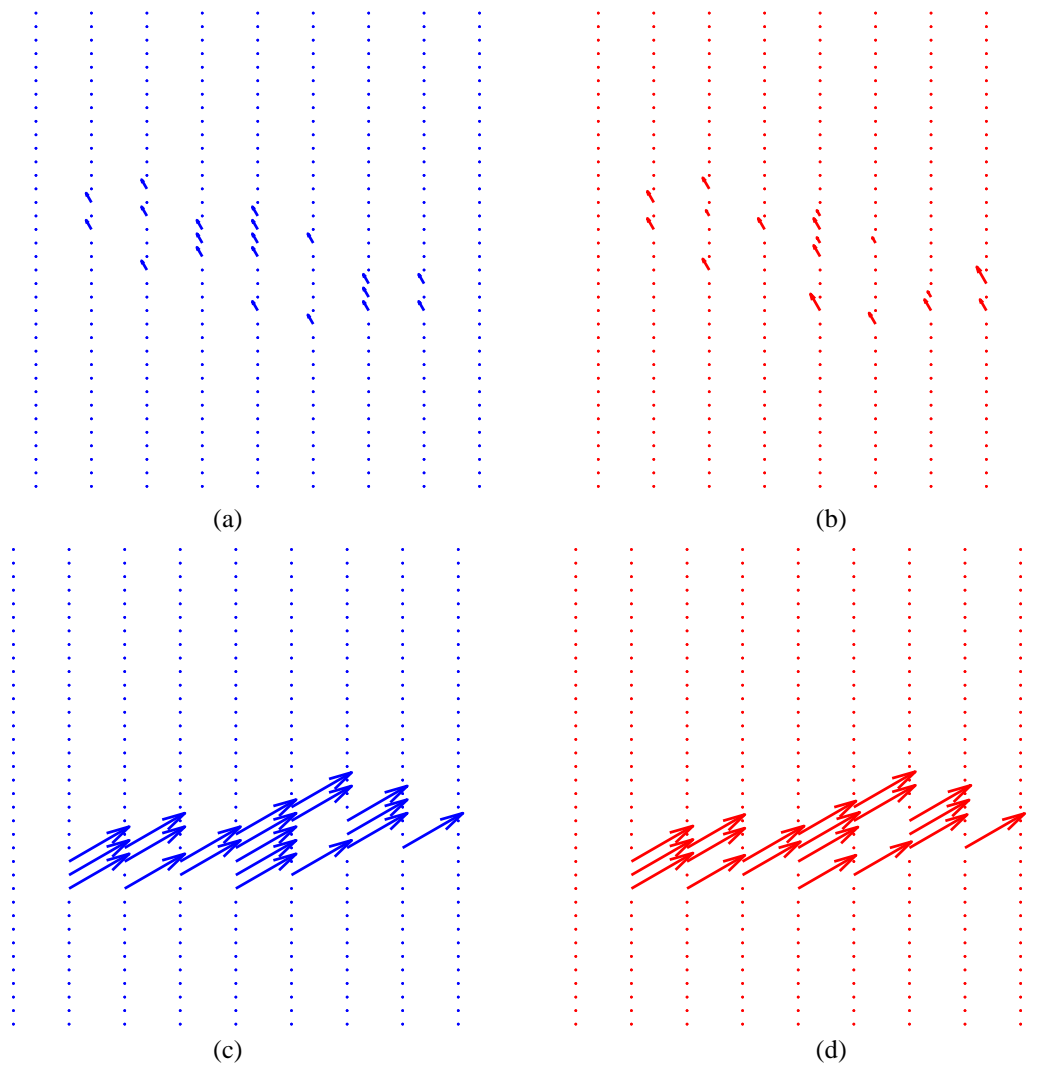

Fig. 6. Velocity estimates with our new, overcomplete dictionary approach for the multi-static, distributed antenna configuration at $SNR = 20$dB. The upper right part of the scene with object moving at 4.7m/s: (a) The true velocity field. (b) The corresponding estimate. The lower left part of the scene with object moving at 32.5m/s: (c) The true velocity field. (d) The corresponding estimate.

| | multi-static, OCD, $M = 400$ | mono-static, OCD, $M = 400$ | multi-static, FBP/MF, $M = 400$ | multi-static, FBP/MF, $M = 12,000$ |
|---|---|---|---|---|
| stationary | 0.0015 | 0.0018 | 0.0452 | 0.0040 |
| dynamic, motion ignored | 0.0104 | 0.0129 | 0.0646 | 0.0093 |
| dynamic | 0.0036 | 0.0090 | 0.1378 | 0.0085 |

TABLE I
PER PIXEL REFLECTIVITY MAGNITUDE ERROR OF DIFFERENT RECONSTRUCTION SCENARIOS WITH A DIFFERENT NUMBER OF MEASUREMENTS $M$.

## VI. CONCLUSION

The radar imaging of scenes that contain motion has long been an interesting and challenging research topic. We have considered multi-static, distributed antenna configurations for high-resolution localization of scenes containing both moving and stationary scatterers. We have presented an overcomplete dictionary inversion approach to simultaneous imaging of stationary and moving scatterers. The non-linear, coupled problem of joint velocity and reflectivity estimation is effectively linearized through introduction of an appropriately defined overcomplete velocity dictionary. The resulting optimization problem is then approximately solved through convex relaxation. Initial experimental results were presented showing the potential of the method for multi-static configurations with narrow band transmissions. In contrast to the existing matched-filtering approaches that are concerned with the inversion of the forward operator only, our overcomplete dictionary formulation explicitly and jointly encodes the sparse point scattering assumption of both the spatial and the velocity dimension, leading to focused imagery with more focused object detail. Initial results suggest the new method exhibits improved robustness to data loss over existing approaches.

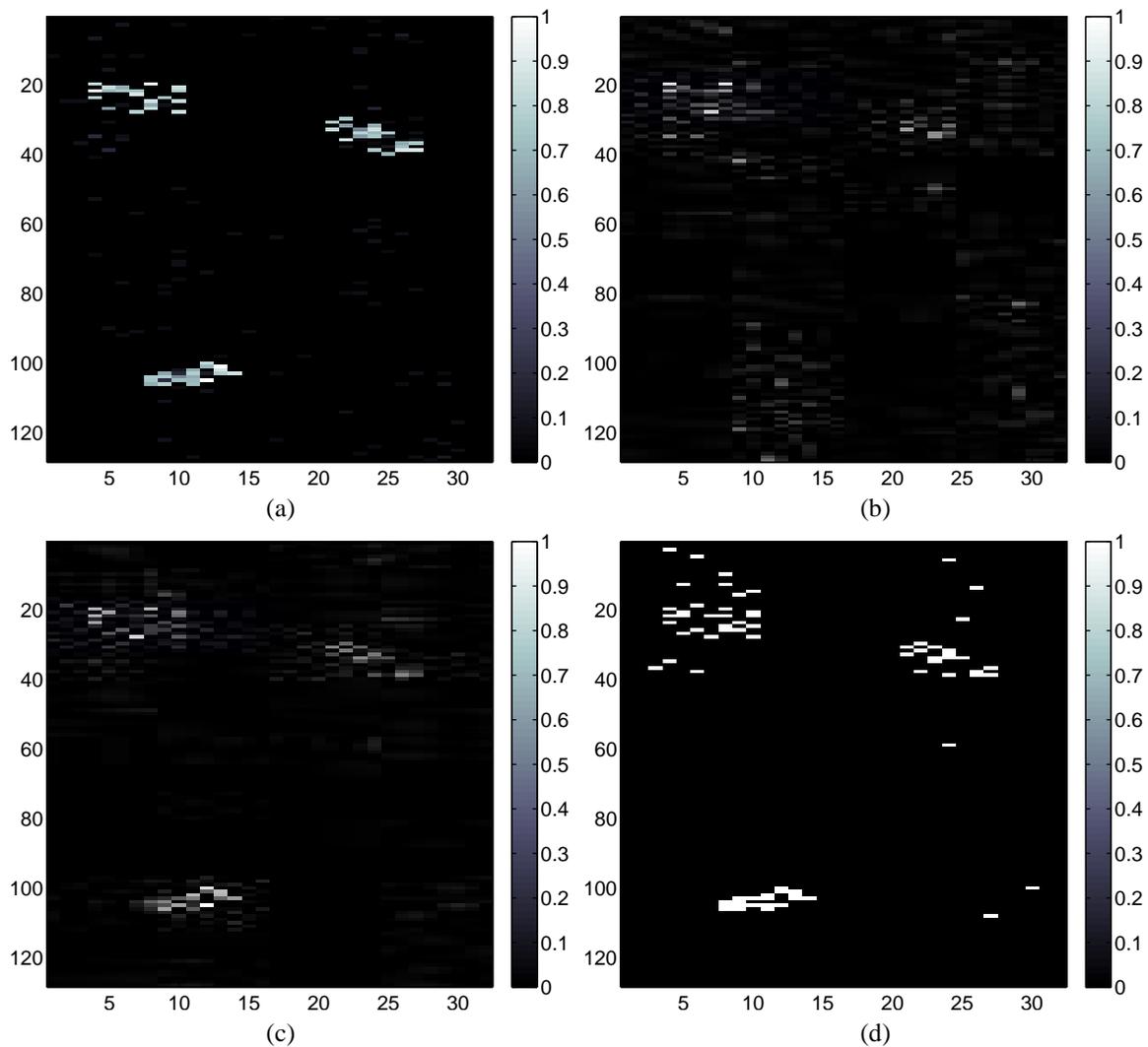

Fig. 7. Reflectivity magnitude reconstruction with our approach for the wide-angle mono-static configuration with $SNR = 20$dB and 400 measurements: (a) The reconstruction of the stationary scene assuming no motion. (b) The reconstruction of the dynamic scene when velocities are ignored, ($\bar{\mathbf{V}} = \{\mathbf{v}_1 = \mathbf{0}\}$). (c) The reconstruction of the dynamic scene with the full overcomplete velocity dictionary. (d) The corresponding locations of reflectors in the reconstruction of (c) whose magnitudes are greater than 0.2.

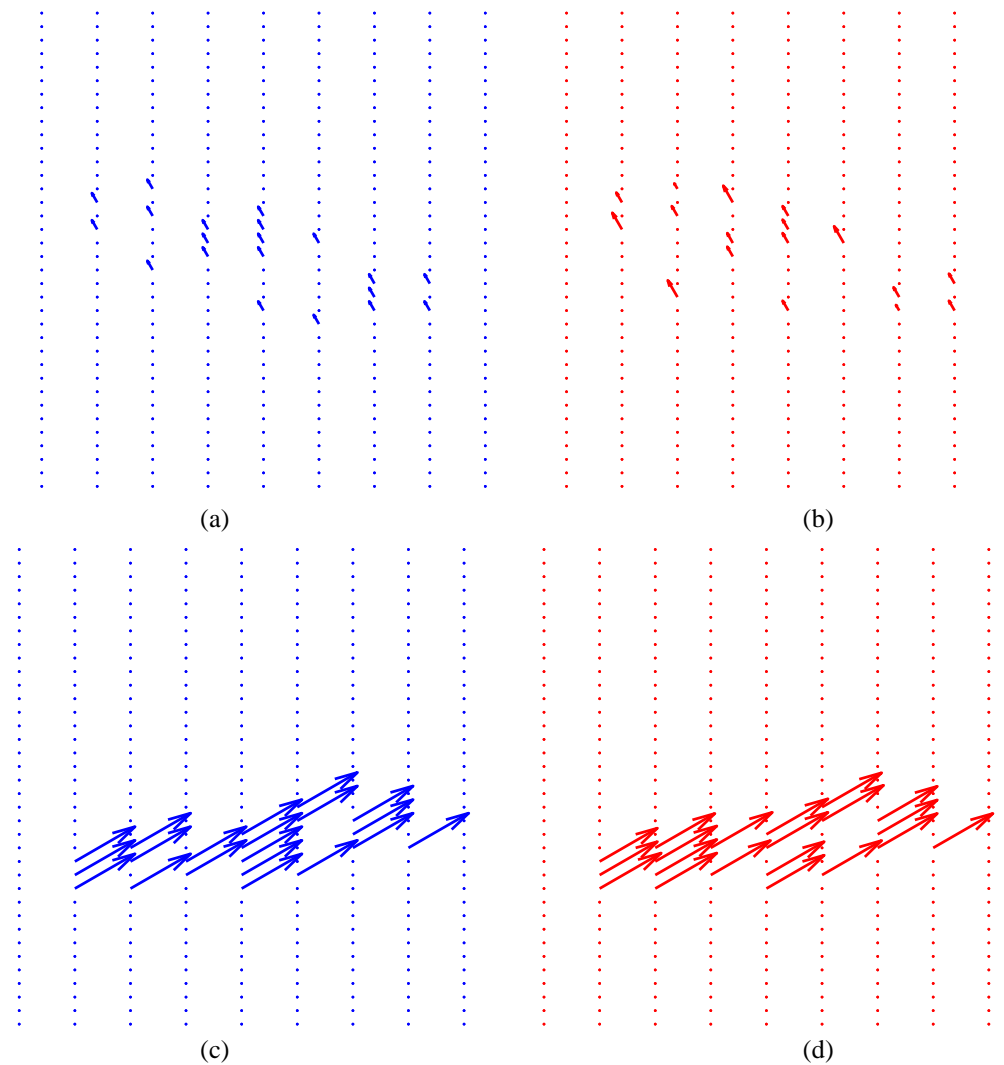

Fig. 8. Velocity estimates with our new, overcomplete dictionary approach for the wide-angle, mono-static configuration at $SNR = 20$dB. The upper right part of the scene with object moving at 4.7m/s: (a) The true velocity field. (b) The corresponding estimate. The lower left part of the scene with object moving at 32.5m/s: (c) The true velocity field. (d) The corresponding estimate.

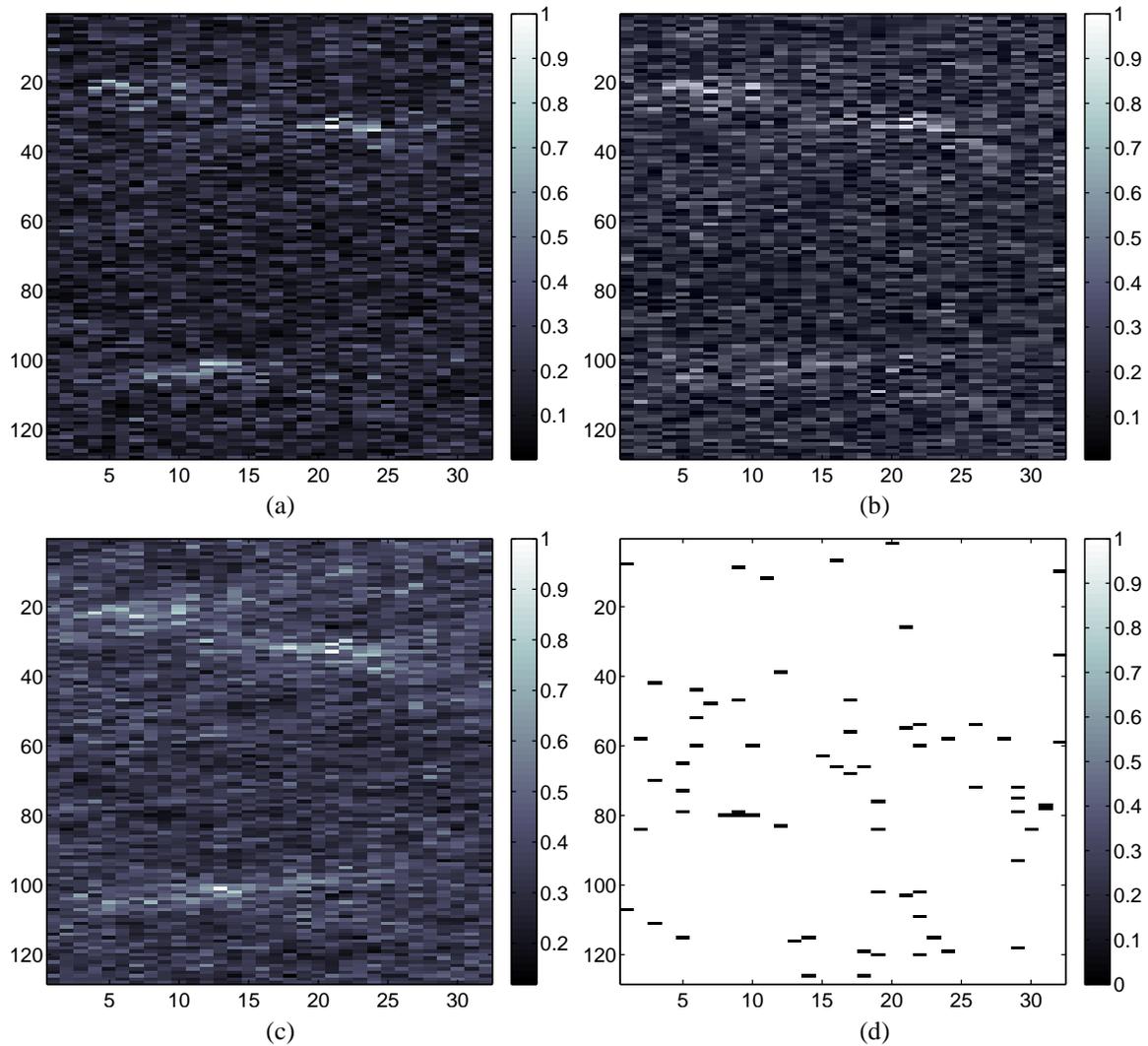

Fig. 9. Reflectivity magnitude reconstruction of the matched-filtering/filtered backprojection approach for the multi-static, distributed antenna configuration at $SNR = 20$dB with 400 measurements, $(N_{tx}, N_{rx}, N_f) = (10, 40, 1)$: (a) The reconstruction of the stationary scene assuming no motion. (b) The reconstruction of the dynamic scene when velocities are ignored. (c) The maximum reflectivity response in the estimated space-velocity cube for the dynamic scene. (d) The corresponding locations of reflectors in the reconstruction of (c) whose magnitudes are greater than 0.2.

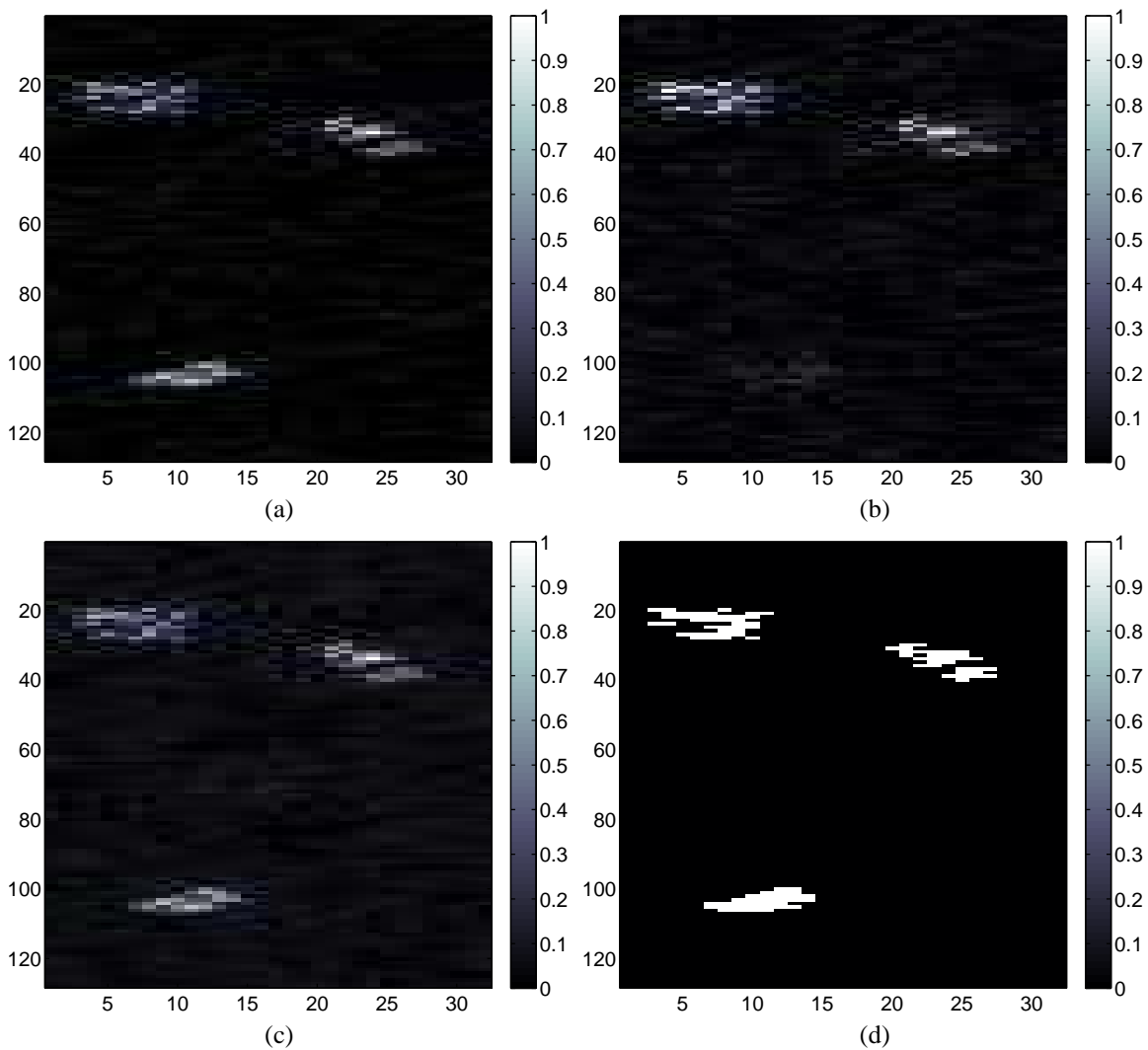

Fig. 10. Reflectivity magnitude reconstruction of the matched-filtering/filtered backprojection approach for the multi-static, distributed antenna configuration at $SNR = 20$dB with $12,000$ measurements, $(N_{tx}, N_{rx}, N_f) = (10, 40, 30)$: (a) The reconstruction of the stationary scene assuming no motion. (b) The reconstruction of the dynamic scene when velocities are ignored. (c) The maximum reflectivity response in the estimated space-velocity cube for the dynamic scene. (d) The corresponding locations of reflectors in the reconstruction of (c) whose magnitudes are greater than 0.2.